\documentclass[prd,
twocolumn,superscriptaddress,preprintnumbers,nofootinbib]{revtex4}
\usepackage{graphicx}
\usepackage{epsfig}
\usepackage{bm}
\usepackage{latexsym,amssymb,amsmath,amssymb,wasysym,float}

\usepackage{color}

\usepackage{enumitem}

\newcommand{\postscript}[2]{\setlength{\epsfxsize}{#2\hsize}
   \centerline{\epsfbox{#1}}}

\usepackage[usenames,dvipsnames]{xcolor}
\definecolor{orange}{cmyk}{0,0.5,1,0}
\definecolor{rossoCP3}{cmyk}{0,.88,.77,.40}
\definecolor{graa}{rgb}{0.8,0.8,0.8}
\definecolor{blaa}{rgb}{0.2,0.2,0.6}

\begin{document}

\preprint{MPP-2024-182}
\preprint{LMU-ASC 15/24}

\title{\color{rossoCP3} Through the Looking Glass into the Dark
  Dimension:\\ Searching for Bulk Black Hole Dark Matter with
 Microlensing of $\bm{X}$-ray Pulsars}

\author{Luis A. Anchordoqui}

\affiliation{Department of Physics and Astronomy,\\  Lehman College, City University of
  New York, NY 10468, USA
}

\affiliation{Department of Physics,\\
 Graduate Center,  City University of
  New York,  NY 10016, USA
}

\affiliation{Department of Astrophysics,
 American Museum of Natural History, NY
 10024, USA
}

\author{Ignatios Antoniadis}

\affiliation{High Energy Physics Research Unit, Faculty of Science, Chulalongkorn University, Bangkok 1030, Thailand}

\affiliation{Laboratoire de Physique Th\'eorique et Hautes \'Energies
  - LPTHE \\
Sorbonne Universit\'e, CNRS, 4 Place Jussieu, 75005 Paris, France
}

\author{Dieter\nolinebreak~L\"ust}

\affiliation{Max--Planck--Institut f\"ur Physik,  
 Werner--Heisenberg--Institut,
80805 M\"unchen, Germany
}

\affiliation{Arnold Sommerfeld Center for Theoretical Physics, 
Ludwig-Maximilians-Universit\"at M\"unchen,
80333 M\"unchen, Germany
}

\author{Karem Pe\~nal\'o Castillo}
\affiliation{Department of Physics and Astronomy,\\  Lehman College, City University of
  New York, NY 10468, USA
}

\begin{abstract}
  \noindent Primordial black holes (PBHs) hidden in the
incredible bulk of the dark dimension could escape constraints from
non-observation of their Hawking radiation. Since these
five-dimensional (5D) PBHs are bigger, colder, and
longer-lived than usual 4D PBHs of the same mass $M$, they could
make all cosmological dark matter if $10^{11} \alt M/{\rm g} \alt
10^{21}$, i.e., extending the 4D allowed region far down  the asteroid-mass window. We show
that these evasive PBHs could be search for by measuring
their $X$-ray microlensing events from faraway pulsars. We also show that
future $X$-ray microlensing experiments will be able to probe the
interesting range ($10^{16.5} \alt M/{\rm g} \alt 10^{17.5}~{\rm g}$) where an all dark matter interpretation in terms of
4D Schwarzschild PBHs is excluded by the non-observation of their Hawking radiation.
\end{abstract}
\date{September 2024}

\maketitle

\section{Introduction}

The dark dimension, an innovative five-dimensional (5D)
scenario that has a compact space with characteristic length-scale in
the micron range, provides a stepping stone to address the
cosmological hierarchy problem~\cite{Montero:2022prj}. This is because the anti-de Sitter
 distance conjecture in de Sitter space~\cite{Lust:2019zwm} connects the size of
the compact space $R_\perp$ to the dark energy scale $\Lambda^{1/4}$
via $R_\perp \sim \lambda \Lambda^{1/4}$, where
$\Lambda \sim 10^{-122} M_p^4$ is the cosmological constant, $M_p$ the Planck mass, and the proportionality factor is estimated to
be within the range $10^{-1} < \lambda <
10^{-4}$~\cite{Montero:2022prj}.

Primordial black holes (PBHs), presumably formed through the collapse
of sufficiently sizable overdensities originated by an enhancement in the comoving curvature perturbation
power spectrum at small scales during inflation~\cite{Zeldovich:1967lct,Hawking:1971ei,Carr:1974nx}, may be one of the
most interesting candidates of dark
matter~\cite{Carr:2020xqk,Green:2020jor,Villanueva-Domingo:2021spv}. These intriguing dark matter candidates are particular
important in the dark dimension scenario, because 5D PBH are bigger, colder, and
longer-lived than usual 4D PBHs of the same mass
$M$~\cite{Anchordoqui:2022txe,Anchordoqui:2022tgp,Anchordoqui:2024dxu}.\footnote{The
  dark dimension scenario provides a specific realization of the dynamical dark
  matter framework~\cite{Dienes:2011ja}
 with other
  interesting dark matter candidates which  have been discussed in~\cite{Gonzalo:2022jac,Anchordoqui:2023tln,Law-Smith:2023czn,Obied:2023clp,Gendler:2024gdo}.} Furthermore, we
have recently shown that a back reaction of Hawking evaporation
process is to kick 5D Schwarzschild PBHs out of the
brane~\cite{Anchordoqui:2024jkn}. As a consequence, 5D PBHs could
make all cosmological dark matter if $10^{11} \alt M/{\rm g} \alt
10^{21}$, i.e., well down into the asteroid-mass window. In this paper
we investigate a potential signal that may allow us to unmask these as
yet evasive 5D PBHs. 

An astonishing coincidence is that $R_\perp$ corresponds approximately to the wavelength of
visible light. This implies that the Schwarzschild radius of black holes perceiving the dark dimension is well below the wavelength of light. For point-like lenses,
this is precisely the critical length where geometric optics breaks
down and the effects of wave optics suppress the magnification~\cite{Matsunaga:2006uc,Sugiyama:2019dgt,Croon:2020ouk},
obstructing the sensitivity to 5D PBH
microlensing
signals. Nevertheless, it was pointed out in~\cite{Bai:2018bej} that
$X$-ray pulsars, with photon energies above 1~keV, are good candidate
sources to search
for microlensing of PBHs with $M \alt 10^{21}~{\rm g}$. In this paper we reexamine this idea focussing attention on 5D PBHs perceiving the
dark dimension. 

The layout is as follows.
In Sec.~\ref{sec:2} we review the generalities of gravitational
microlensing. In Sec.~\ref{sec:3} we focus attention on the
particulars of microlensing of $X$-ray pulsars, which is relevant to unmask PBHs perceiving
the dark dimension. The
paper wraps up in Sec.~\ref{sec:4} with some conclusions.

\section{Point-lens Microlensing Basics}
\label{sec:2}

\begin{figure}[tpb]
\postscript{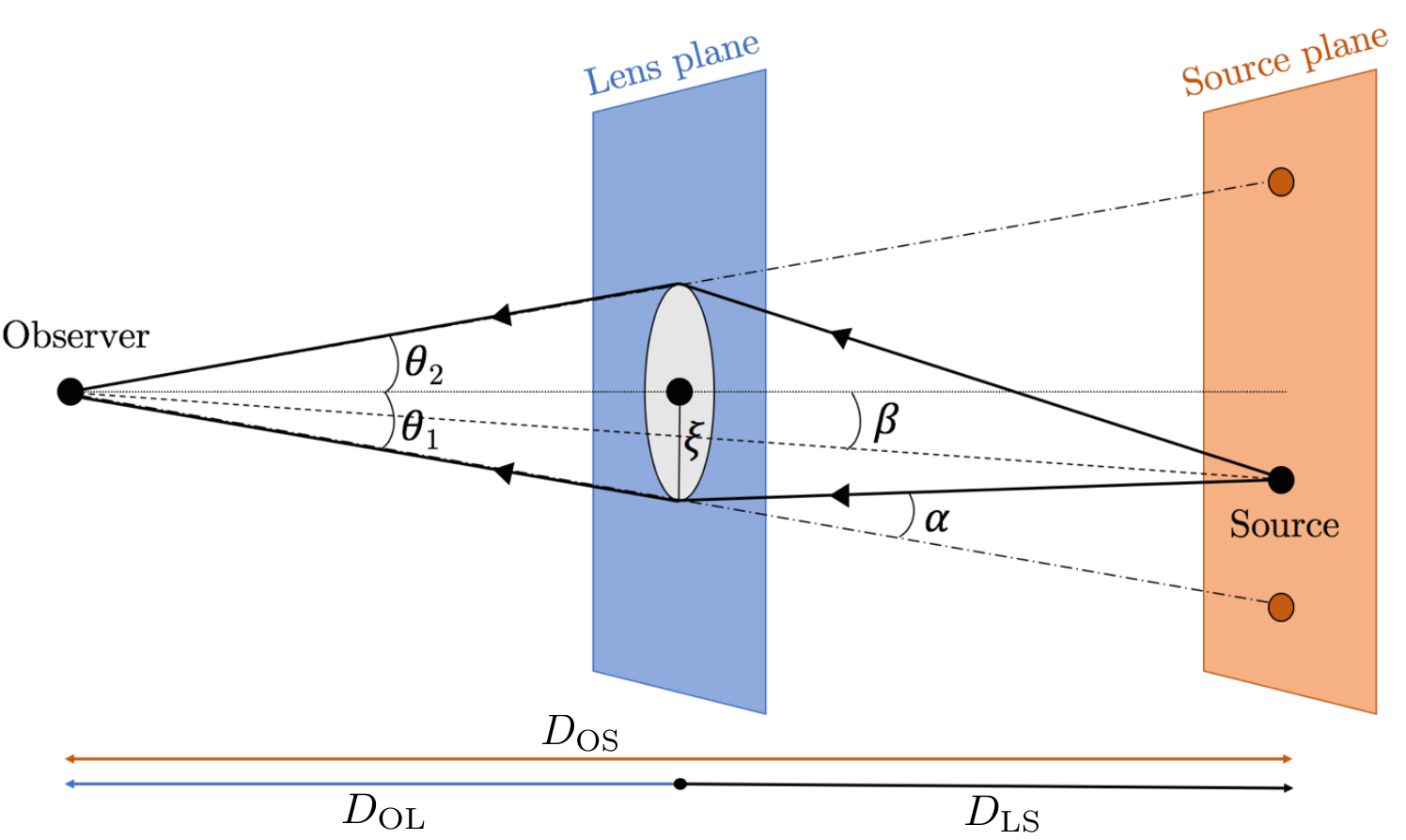}{0.95}
\caption{The geometry of the microlensing setup for a point-like lens,
  seen to produce two images after deflection~\cite{Croon:2020wpr}. \label{fig:1}}
\end{figure}

It has long been known that gravitational lensing becomes observable when
a massive object is located between a light-emitting source and an
observer~\cite{Einstein:1936llh}. This is because the gravitational potential of the central object acts as a lens that bends
the path of the light rays coming from the source, by bending the
spacetime through which the photon travels. If the central object is a Schwarzschild
black hole, described by the line element
\begin{equation}
  ds^2 = - f(r) \ dt^2 + f^{-1}(r) \ dr^2 + r^2 \ d\Omega_2^2 \,,
\label{ds4}
\end{equation}
the deflection
angle of a light ray is estimated to be
\begin{equation}
  \alpha = \frac{4GM}{r_0} + 4 \left(\frac{GM}{r_0}\right)^2  \left(\frac{15
      \pi}{16} - 1 \right) + {\cal O} \left(\epsilon^3 \right) \,,
  \label{alpha}
\end{equation}  
where $f(r) = 1 -2GM/r$ is the blackening function, $M$ is the mass of
the black hole (i.e. the lens), $G =M_p^{-2}$ is Newton's
gravitational constant, $d\Omega_2^2$ is the metric of a 2-dimensional unit sphere, and $r_0$ is the closest distance of approach to
the lens in the lens-Earth-source plane, with $\epsilon = GM/r_0$, and
where all measurements are taken in the observer's reference
frame~\cite{Virbhadra:1999nm}. To develop some sense for the orders of magnitude involved,
we consider the set up shown in Fig.~\ref{fig:1} for a point-like
lens. In the absence of the lens, we would
then observe the source at an angular position $\beta$ on the source
plane, but because of the deflection, we actually observe it at
$\theta$. Since the observer distances to the lens $D_{\rm OL}$ and source
$D_{\rm OS}$ are very large in
relation to the deflection angle, the distances of the source and
image from the optical axis in the source plane are estimated using the small angle approximation
\begin{equation}
  \theta D_{\rm OS} = \beta D_{\rm OS} + \alpha D_{\rm LS} \,,
\label{lenseq}
\end{equation}
where $D_{\rm LS}$ is the distance from the lens to the source~\cite{Matsunaga:2006uc}.

The impact parameter of the unperturbed light ray is found to be
$\xi = r_0 (1-2GM/r_0)^{-1/2
}$. In the point-lens approximation, however, $\xi \equiv
\theta D_{\rm OL} \sim r_0$,
and so (\ref{lenseq}) can be rewritten as
\begin{equation}
  \theta^2 = \beta \theta + \theta_E^2 \,,
\label{lenseq2}
\end{equation}  
where
\begin{equation}
  \theta_E = \sqrt{4GM \ \frac{D_{\rm LS}}{D_{\rm OL} \ D_{\rm OS}}}
\end{equation}  
is the angular size of the Einstein ring which is formed when the source is
perfectly aligned behind the lens, i.e., when $\beta = 0$, and where
we have neglected terms ${\cal O}(\epsilon^2)$. $\theta_E$ in turn defines the
Einstein radius $r_E = D_{\rm OL} \theta_E$ on the lens
plane. Note that for $\beta \sim 0$, the impact parameter
  $\xi \sim r_E$ is the characteristic   scale
  of the source-lens-observer system. All in all,
the isolated Schwarzschild black hole will split a point source into
two images with angular positions defined implicitly by
\begin{equation}
 \theta_{1,2} = \frac{1}{2} \left(\beta \pm \sqrt{\beta^2 + 4
     \theta_E^2} \right) \, .
\end{equation}
The two images are on opposite sides of the source, with one image inside the Einstein ring and the other outside. As the source moves away from the lens (i.e. as $\beta$ increases), one of the images approaches the lens and becomes very faint, whereas the other image approaches the true position of the source and asymptotes to its unlensed flux.

A point worth noting at this juncture is that if a source is closer than $\theta_E$ in separation from a lensing PBH
on the sky, the source is multiply imaged by its
lensing. Nevertheless, the separation between multiple images is too small
to be resolved by optical telescopes. What is observed instead in the so-called
``microlensing phenomenon'' is the temporary magnification of the total flux of two images relative to that of the
unlensed source. Now, if a gravitational field were to deflect a
light ray this would produce a change in the cross-section of a bundle
of rays. However, according to Liouville's theorem the phase space
density of photons must be conserved, which implies that gravitational lensing
should preserve the surface brightness of the source and should only
change its apparent surface area. In other words, the magnification of
an image becomes the ratio of the solid angles of the image
and of the unlensed source (at the observer position). If the central
object is spherically symmetric, the magnification factor is found to be
\begin{equation}
  \mu = \frac{\sin \theta \ d \theta}{\sin \beta \ d \beta} \simeq
  \frac{\theta}{\beta} \ \frac{d \theta}{d\beta} \, ,
\label{mueq}  
\end{equation}
where the sign of the magnification gives the parity of the particular
image. Substituting $\beta$ from the point-lens equation (\ref{lenseq2})
into (\ref{mueq}), it follows the magnifications of the two images,
   \begin{equation}
     \mu_{1,2} = \left[1- \left(\frac{\theta_E}{\theta_{1,2}}\right)^4
       \right]^{-1} = \frac{u^2 +2}{2 u \sqrt{u^2 + 4}} \pm \frac{1}{2} \,,
   \end{equation}
where $u$ is the angular separation of the source from the point mass
in units of the Einstein angle, $u = \beta/\theta_E$. Note that $\theta_2 <
\theta_E$ implies $\mu_2 < 0$, and so the magnification of the image which is inside the Einstein ring is negative implying that this image has its parity flipped with respect to the source. The net magnification of flux in the two images is obtained by adding the absolute magnifications,
\begin{equation}
  \mu_{\rm tot} (u) = \mu_1 + \mu_2 = \frac{u^2 +2}{u \sqrt{u^2 +4}} \, .
\end{equation}
Note that when the source lies on the Einstein radius, we have $\beta =
\theta_E$ and u = 1, so that the total magnification becomes $\mu_{\rm
  tot} = 1.17 + 0.17 = 1.34$.

Now, the source and the lens have a relative motion with respect to
the observer. As a consequence, the observed flux of a source varies
with time, yielding a characteristic light curve of the observed
source flux. Such a unique light curve allows a microlensing event to be identified from
the observation of other variable sources. A typical timescale of the
microlensing light curve can be estimated from a crossing time of the
Einstein radius for a lensing PBH with respect to a distant source,
\begin{equation}
t_E = r_E/v \,, 
\end{equation}
where $v$ is the relative velocity for a observer-lens-source system~\cite{Niikura:2017zjd}.

Microlensing surveys are typically sensitive to stars that are $10 \alt
 D_{\rm OS}/{\rm kpc} \alt 1000$ away and to transit times of minutes
 to years. For example, the Subaru-HSC instrument was sensitive to the
 short transit times of light PBHs reaching sensitivities to constrain
 an all dark matter 
 PBH interpretation for masses $M \agt 10^{22}~{\rm g}$~\cite{Matsunaga:2006uc,Sugiyama:2019dgt,Croon:2020ouk}. However the Subaru-HSC survey was
 limited by two effects as the PBH mass is reduced:
\begin{itemize}[noitemsep,topsep=0pt]
\item the finite apparent size of the
source stars in its sky target, M31, being larger than the apparent
Einstein radius, yielding poorly focused light;
\item transition
from geometric to wave optics as the PBH Schwarzschild radius becomes
comparable to or smaller than the wavelength of optical light.
\end{itemize}
Indeed, the wave effects characterized by 
\begin{equation}
 w =  \frac{8 \pi G M}{\lambda} = 0.3 \left(\frac{M}{10^{22}~{\rm g}} \right)
   \left(\frac{\lambda_0}{621~{\rm nm}}\right)^{-1} \,,
\label{w}   
\end{equation}    
become important when $w \alt 1$, where $\lambda_0$ is the characteristic wavelength of light in an
observation~\cite{Gould:1992}; the default choice in the normalization of (\ref{w})
corresponds to a central wavelength of $r$-band in the Subaru
telescope. As it was first pointed out in~\cite{Bai:2018bej},
consideration of sources emitting in the $X$-ray spectrum could allow
us to search deep into the PBH low-mass window. It is this that we now turn to study.

\section{Microlensing of black holes perceiving the dark dimension}
 \label{sec:3}

It is a known fact that $M_p$ is the natural cutoff of quantum
gravity. However, if there were light species of particles in the
theory, then the consistency of black hole entropy with the effective field
theory description would demand a breakdown of the
classical picture at a lower scale,
\begin{equation}
M_* =  m_{\rm KK}^{(d-4)/(d-2)}    M_p^{2/(d-2)} \,,
\end{equation}  
dubbed the species
scale, where $d$ is the spacetime dimension and $m_{\rm KK} \sim R_\perp^{-1}$~\cite{Dvali:2007hz,Dvali:2007wp}. Now, despite
the fact that $M_*$ is motivated by the emergence of the tower of
light states, curiously  
the mass scale of this tower, $m_{\rm KK}$, does not seem to be directly captured by $M_*$~\cite{Bedroya:2024uva}.

One way the lower-dimensional theory can find out about the $m_{\rm KK}$ scale is
through the study of black holes~\cite{Anchordoqui:2022tgp,Anchordoqui:2024dxu,Bedroya:2024uva}. When black holes get smaller than
$R_\perp$, the black hole becomes thermodynamically unstable because
of the Gregory–Laflamme
transition~\cite{Gregory:1993vy}. Black holes
with $r_s \gg R_\perp$ are actually black branes wrapped around the extra dimensions, but
the ones with $r_s \ll R_\perp$ are localized in the extra dimensions. This
transition to a new black hole solution marks a new scale $\Lambda_{\rm BH}$
in the lower-dimensional theory~\cite{Bedroya:2024uva}. In this section we show that Earth-based microlensing
experiments are insensitive to the $\Lambda_{\rm BH}$ scale, independently of the black hole mass $M$.

Throughout we assume that the Standard Model fields are confined to a
D-brane and only gravity spills into the compact space of dimension $(d-4)$~\cite{Antoniadis:1998ig}.

\subsection{Microlensing of Bulk Black Holes}

The Gregory-Laflamme transition induces a change in the scaling of the black hole's Schwarzschild radius $r_s$ and its Hawking temperature $T_H$~\cite{Anchordoqui:2022tgp}. If the black hole is spherically
symmetric and $r_s \ll R_\perp$, then it can be treated as a flat
$d$-dimensional object with line element given by
\begin{equation}
ds^2 = -U(r) \ dt^2 + U^{-1} (r) \ dr^2 + r^{2} \ d\Omega_{d-2}^2\,,
\end{equation}
where
$U(r) = 1 - (r_s/r)^{d-3}$
is the blackening function,
\begin{equation}
  d\Omega^2_{d-2} = d \chi_2^2 + \prod_{i=2}^{d-2} \sin^2 \chi_i \
  d\chi^2_{i+1}
\end{equation}  
is the metric of a $(d-2)$-dimensional unit sphere, and
\begin{equation}
    r_s = \frac{1}{M_*} \left[ \frac{M}{M_*} \right]^{1/(d-3)}
    \left[ \frac{8 \, \Gamma\big(\frac{d-1}{2} \ \big)}{(d-2) \ \pi^{(d-3)/2} }
    \right]^{1/(d-3)} \,, 
\label{r_s}
  \end{equation}
and where $\Gamma(x)$ is the Gamma
function~\cite{Tangherlini:1963bw,Myers:1986un,Argyres:1998qn}. The 
$d$-dimensional black hole behaves like a thermodynamic system~\cite{Hawking:1976de}, with
temperature $T_H \sim (d-3)/(4 \pi r_s)$ and entropy $S = (4 \pi
M r_s)/(d-2)$~\cite{Anchordoqui:2001cg}. If the black hole is
localized on the brane, Hawking evaporation~\cite{Hawking:1974rv,Hawking:1975vcx} proceeds dominantly
through emission of Standard Model
fields~\cite{Emparan:2000rs}. However, we have recently shown that the recoil effect due to graviton emission imparts the black hole a relative kick
velocity with respect to the brane, allowing Schwarzschild black holes to escape into
the bulk~\cite{Anchordoqui:2024jkn}.  Because the escape from the
brane is almost instantaneous 5D Schwarzschild
black holes evaporate almost entirely into gravitons in the
bulk, and so can evade constraints from the non-observation of Hawking
radiation in: {\it (i)}~the extragalactic
$\gamma$-ray background~\cite{Carr:2009jm}, {\it (ii)}~the cosmic microwave
background~\cite{Clark:2016nst}, {\it (iii)}~the 511~keV $\gamma$-ray
line~\cite{DeRocco:2019fjq,Laha:2019ssq,Dasgupta:2019cae,Keith:2021guq},
{\it (iv)} the EDGES 21-cm signal~\cite{Mittal:2021egv,Saha:2021pqf},
{\it (v)}~Lyman-$\alpha$ forest~\cite{Saha:2024ies}, and {\it (vi)}~the MeV Galactic diffuse emission~\cite{Laha:2020ivk,Berteaud:2022tws,Korwar:2023kpy}.

The deflection angle of a light ray expected to be induced  by a $d$-dimensional Schwarzschild
black hole localized on the brane is estimated to be
\begin{equation}
\alpha_d  =  \frac{2 (d-2) M}{M_*^{d-2} \ r_0^{d-3}} \ _2F_1
               \left(\frac{1}{2}, \kappa; \frac{3}{2} ;1 \right)  +  {\cal
  O}\left(\frac{M^2}{M_*^{d} r_0^{d-2}}\right)  \,,
\label{alphad}
\end{equation}
where $_{2}F_{1}(a,b,;c;z)$ is the Gaussian 
hypergeometric function, with
$\kappa = 1/2 - (d-3)/2$~\cite{Lobos:2024fzj}. A straightforward
calculation shows that for $d=4$, (\ref{alphad})
reduces to the first term in (\ref{alpha}). For $d=5$,
\begin{equation}
  \alpha_5 = \frac{3 }{2} \ \frac{\pi}{M_*^3} \ \frac{M}{r_0^2} + {\cal O}
  \left(\frac{M^2}{M_*^5r_0^3}\right) \,, 
\end{equation}    
which implies that $\alpha_5 < \alpha_4$ for fixed $M$ and $r_0$, with $r_s < r_0
< R_\perp$. This was interpreted in~\cite{Lobos:2024fzj} as the need for advanced sensitive detection devices to
observe lensed images influenced by the dark dimension.

The previous statement should be revised with caution. It
  is clear that while the source is on the brane, in the dark dimension scenario the lens has an extension into the
    bulk. Therefore, the light emitted by the source could be a micron
    away from the lens 4D position. The scale that counts in addressing whether the 
5D geometry of the lens would bring important corrections to the source magnification $\mu_{\rm
  tot}$ is actually 
    the size $\xi$ of the lens involved in the deflection. As we have
    stressed 
    in Sec.~\ref{sec:2}, in the point-lens
    approximation of Fig.~\ref{fig:1}, the characteristic scale of the
    source-lens-observer system is
    $\xi \agt r_E$.

  At this point a reality check is in order. A source
    could be a good object to undergo microlensing if it is a long
    distance away from the
  telescopes (viz., the Earth), because this would increase the optical depth and/or the number of possible lensing events. Following~\cite{Bai:2018bej},  we consider the $X$-ray
pulsar SMC X-1 in the Small Magellanic Cloud at a distance $D_{\rm OS}
= 65~{\rm kpc}$~\cite{Hickox:2004fy}. For such a distance, the Einstein radius of a
$M \sim 10^{17}~{\rm g}$ black hole is given by
\begin{eqnarray}
  r_E & = & \sqrt{4GM x (1-x) D_{\rm OS}} \nonumber \\
   & = & 12~{\rm km} \left[\left(\frac{M}{10^{17}~{\rm g}}\right) \left(\frac{D_{\rm
         OS}}{65~{\rm kpc}}\right) \left(\frac{x (1-x)}{1/4}
         \right) \right]^{1/2}
\end{eqnarray}         
where $x = D_{\rm OL}/D_{\rm OS}$. Since $r_E\gg R_\perp$ is the closest
distance that source photons get to the lens when it lies directly
along the line of sight, we conclude that corrections to $\mu_{\rm
  tot}$ due to the 5D
geometry of the lens can be safely neglected. In other words, for $r \gg R_\perp$ the gravitational
potential of a 5D black hole falls as $1/r$ and therefore is indistinguishable from that of a 4D
black hole of the same mass. This implies that for
Earth-based experiments the microlensing signal
of a 5D black hole would be indistinguishable from that of a 4D black
hole of the same mass, and this is independent on whether the black
hole is localized on the brane or propagates through the bulk.

Since line-of-sight distances of interest are usually much larger than
transverse distances, $X$-ray microlensing events can be pictured as
projections on the lens-containing
transverse plane, {\it viz.} the
lens plane. The source radius in the lens plane is estimated to be
\begin{eqnarray}
  a_s(x) & = & \frac{x R_s}{r_E (x)} \sim 0.8 \times \left(\frac{x}{\sqrt{x(1-x)}} \right)
     \left(\frac{R_S}{20~{\rm km}}\right) \nonumber \\
     & \times & \left(\frac{D_{\rm
             OS}}{65~{\rm kpc}} \right)^{-1} \left(\frac{M}{10^{17}~{\rm g}}\right)^{-1/2} \,,
\label{as}
\end{eqnarray}
where following~\cite{Bai:2018bej} we have taken a fiducial source
size $10 \alt R_S/{\rm km} \alt 20$. From (\ref{as}) we infer that $X$-ray
pulsars could overcome finite source size effects for $M \sim
10^{17}~{\rm g}$ if $x \alt 0.6$.  

Now, wave effects becomes important when $w \alt 1$~\cite{Gould:1992} or
equivalently for photon energies
\begin{equation}
E_\gamma \alt \frac{1}{4GM} = 660~{\rm keV}
\left(\frac{M}{10^{17}~{\rm g}}\right)^{-1} \, .
\end{equation}  
Thus, for a source emitting primarily with $X$-ray energy of $1 \alt
E_\gamma/{\rm keV} \alt 10$~\cite{Hickox:2004fy}, wave effects must be
taken into account in order to probe a lower PBH mass of about
$10^{17} ~{\rm g}$.

The magnification, including the wave optics effect, is given by
\begin{equation}
\mu_{\rm tot} (w,u) = \frac{\pi w}{1 - e^{-\pi w}} \ \left| _{1}F_{1}
  \left(\frac{i}{2} w,1;\frac{i}{2} w u^2\right) \right|^2
\end{equation}
where $_{1}F_{1}(a;b;z)$ is the confluent hypergeometric function~\cite{Takahashi:2003ix}. For $u=0$
the magnification has a maximum
\begin{equation}
  \mu_{\rm tot}(w,u)\bigg|_{\rm max} = \mu_w(w,0) = \frac{\pi w}{1 -
   e^{-\pi w}} \, .
\label{muw}
\end{equation}
It is easily seen from (\ref{muw}) that for microlensing events with
$w \ll 1$, the maximum magnification
\begin{equation}
  \mu_{\rm tot} (w,u) \big|_{\rm max} = 1 + \frac{\pi w}{2} \,,
\end{equation}
 would be significantly reduced as compared to the maximum
 magnification $\mu_{\rm tot} (u)
 \to \infty$
in the geometric optics approximation.\footnote{Note that $\mu
_{\rm tot}  (u,w) =1$ means no lensing magnification.} This is because the
gravitational potential induced by extremely light PBHs is too weak to
bent the path of the photons. 

\begin{figure}[tpb]
\postscript{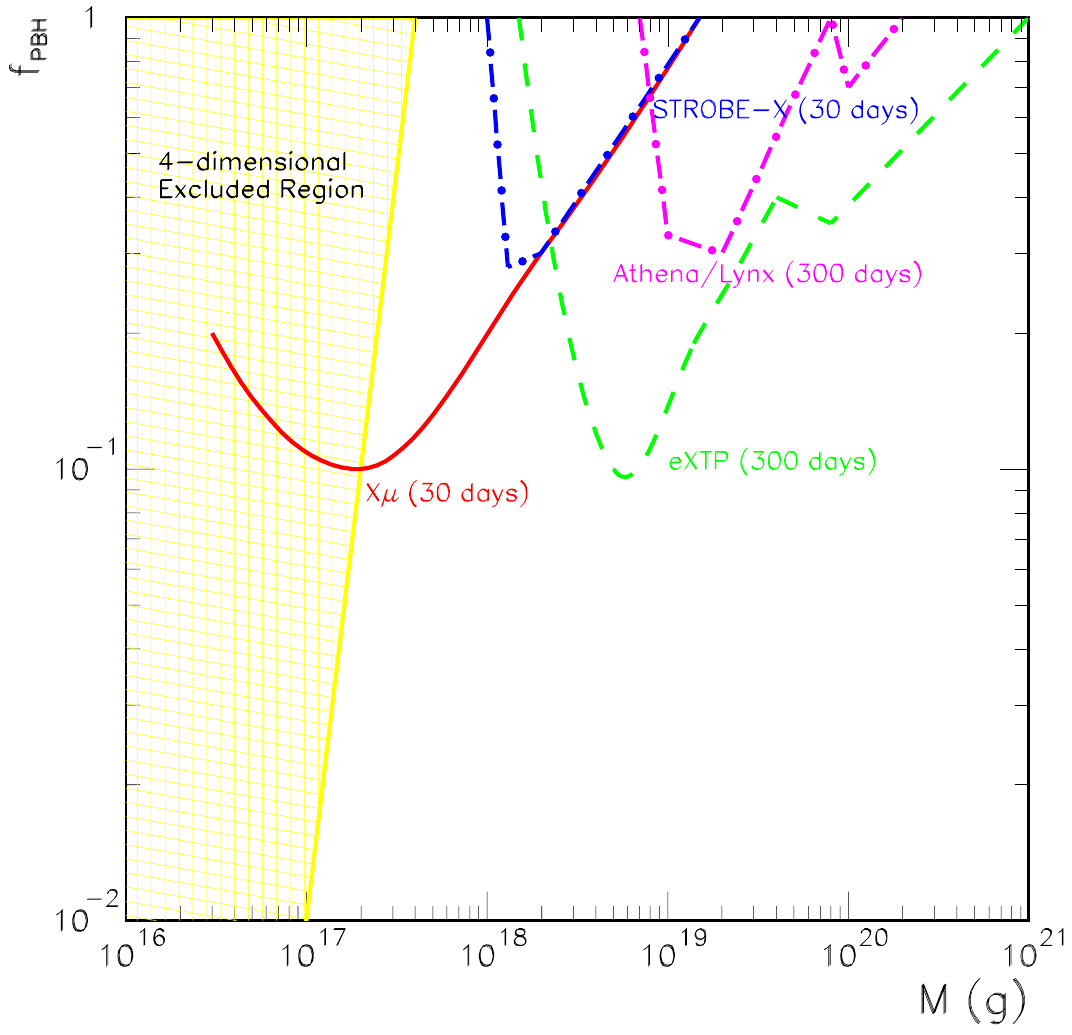}{0.95}
\caption{Compilation of projected sensitivities on the PBH dark matter fraction
  $f_{\rm PBH}$ as a function of $M$ for future $X$-ray
  telescopes,  assuming
  a monochromatic mass function and including wave optics effects. The sensitivity curves have been
  taken from~\cite{Bai:2018bej,Tamta:2024pow}. For comparison, the shaded region
  shows contraints on the 4D $f_{\rm PBH}$ from the non-observation of
  Hawking radiation in: {\it (i)}~the extragalactic
$\gamma$-ray background~\cite{Carr:2009jm}, {\it (ii)}~the cosmic microwave
background~\cite{Clark:2016nst}, {\it (iii)}~the 511~keV $\gamma$-ray
line~\cite{DeRocco:2019fjq,Laha:2019ssq,Dasgupta:2019cae,Keith:2021guq},
{\it (iv)} the EDGES 21-cm signal~\cite{Mittal:2021egv,Saha:2021pqf},
{\it (v)}~Lyman-$\alpha$ forest~\cite{Saha:2024ies}, and {\it (vi)}~the MeV Galactic diffuse emission~\cite{Laha:2020ivk,Berteaud:2022tws,Korwar:2023kpy}. \label{fig:2}.}
\end{figure}

The mass distribution of PBHs is usually characterized by the mass function
\begin{equation}
\psi (M) = \frac{M}{\rho_{\rm CDM}} \ \frac{dn_{\rm
    PBH}}{dM} \,, 
\label{psiM}
\end{equation}
where $dn_{\rm PBH}$ is the number density of PBHs within the mass
range $(M, M + dM)$, and $\rho_{\rm CDM}$
is the energy density of cold dark matter~\cite{DeLuca:2020ioi}. Integrating $\psi (M)$ gives the total fraction of dark matter in PBHs,
\begin{equation}
f_{\rm PBH} \equiv \frac{\rho_{\rm PBH}}{\rho_{\rm CDM}} = \int \psi
(M) \, dM \,, 
\end{equation}
where $\rho_{\rm PBH} = \int M \, dn_{\rm PBH}$ is the energy
density of PBHs. If all of the dark matter were made of PBH,
we would have $f_{\rm PBH} = 1$. In Fig.~\ref{fig:2} we show a compilation of the projected sensitivity to $f_{\rm PBH}$ of future
$X$-ray telescopes, including Athena~\cite{Barcons:2015dua}, Lynx~\cite{LynxTeam:2018usc}, and eXTP for a
300-day observation~\cite{eXTP:2016rzs},
as well as STROBE-X~\cite{STROBE-XScienceWorkingGroup:2019cyd}, and $X\mu$ for
a 30-day observation~\cite{Tamta:2024pow}. We can see that these
instruments will break down into the mass range relevant for bulk
black hole dark matter, including the interesting region in which a
4D PBH all dark matter interpretation has been excluded due to the
non-observation of Hawking radiation.

\subsection{Microlensing of Near-Extremal Black Holes }

According to the no-hair theorem~\cite{Ruffini:1971bza}, 4D black hole geometries in
asymptotically flat spacetimes eventually settled down to
Kerr-Newman solutions~\cite{Kerr:1963ud,Newman:1965}, which are characterized by three measurable parameters: the mass $M$, the
angular momentum $\vec J$ and the electric charge $Q$. An important feature of
Kerr-Newman black holes is that the three
parameters are
not all independent from each other. Actually, for a given set of parameters there is a minimal extremal mass $M_{\rm e}$ satisfying
\begin{equation}
M^2 \geq M_{\rm e}^2 =  \left(\frac{M_p^2J}{M}\right)^2 + (M_p Q)^2 \, ,
\end{equation}
where $Q$ is measured in units in which the Coulomb force between two
charges separated by a distance $d$ has magnitude
$F=Q^2/d^2$ and  $J = |\vec J|$~\cite{Wald:1974hkz}. If the parameters saturate this bound the black hole is dubbed
extremal, and if the parameters are close to saturating it
near-extremal. Before proceeding we pause to note that:
\begin{itemize}[noitemsep,topsep=0pt]
\item When the black hole mass is tuned below $M_{\rm ext}$,
the event horizon disappears leaving behind a naked singularity, which
violates the cosmic censorship conjecture~\cite{Penrose:1964wq}.
\item When a black hole reaches its extreme limit, the thermal
description breaks down, and it cannot continue to evaporate by emitting
(uncharged) elementary particles.
\item When black holes are near-extremal their evaporation temperature
  decreases, and consequently so does their luminosity. Thus,
  near-extremal black holes can also evade constraints from the non-observation of Hawking
radiation~\cite{deFreitasPacheco:2023hpb,Anchordoqui:2024akj}.
\end{itemize}

It has long been suspected that any electromagnetic charge or
spin would be lost very quickly by any 4D black hole
population of 
primordial origin. On the one hand, the electromagnetic charge of a black hole is spoiled by the Schwinger
effect~\cite{Schwinger:1951nm}, which allows pair-production of
electron-positron pairs in the strong electric field outside the
black hole, leading to the discharge of the black hole and subsequent
evaporation~\cite{Gibbons:1975kk,Hiscock:1990ex}. On the other hand, a rapidly
rotating black hole spins down to a nearly non-rotating state
before most of its mass has been given up, and therefore it does not
approach to extremal when it evaporates~\cite{Page:1976ki}. All in
all, near-extremal primordial Kerr-Newman black holes  are
not expected to prevail in the universe we live in.

An alternative
interesting possibility is to envision a scenario where the black hole
is charged under a generic unbroken $U(1)$ symmetry (dark photon),
whose carriers (dark electrons with a mass $m'_e$ and a gauge coupling
$e'$) are always much heavier than the temperature of the black
hole~\cite{Bai:2019zcd}. This implies that the charge $Q$ does not get
evaporated away from the black hole and remains therefore
constant. Strictly speaking, the pair production rate per unit volume
from the Schwinger effect can be slowed down by arbitrarily decreasing
$e'$, whereas the weak gravity conjecture (WGC) imposes a constraint on the
charge per unit mass; namely, for each conserved gauge charge there
must be a sufficiently light charge carrier such that
\begin{equation}
e'q/m_{e'} \geq
\sqrt{4\pi} \sqrt{(d-3)/(d-2)} \ \bar M_p^{-(d-2)/2} \,,
\label{dWGC}
\end{equation}
where $q$ is the
integer-quantized electric charge of the
particle and $\bar M_p = M_p/(8\pi)$ is the reduced Planck mass~\cite{Arkani-Hamed:2006emk,Harlow:2022ich}. Setting $e = e' = \sqrt{4 \pi
  \alpha}$ the (4D) Schwinger effect together with the WGC lead to a bound on the minimum black hole mass of near
extremal black holes with evaporation time longer than the age of
Universe, $M_{\rm ne}  \gtrsim 5 \times 10^{15}~{\rm g} (m_{e'}/10^9~{\rm
  GeV})^{-2}$~\cite{Bai:2019zcd}.

Since we have seen that microlensing experiments cannot distinguish 4D
from 5D black holes, in what follows we consider the line element (\ref{ds4}) with
blackening function
\begin{equation}
f(r) = 1 - \frac{2M}{M_p^2 \ r} + \frac{Q^2}{M_p^2 \ r^2} \, .
\label{RN}
\end{equation}
The deflection angle of a light ray expected to be induced by the
geometry given in (\ref{RN}) has been computed in~\cite{Eiroa:2002mk} and is given by 
\begin{equation}
  \alpha_Q = \frac{4 M}{M_p^2 \ r_0} + \left(\frac{15}{16} \pi -
      1 \right) \frac{4 M^2}{M_p^4 r_0^2} - \frac{3}{4} \pi
    \frac{Q^2}{M_p^2 r_0^2} + \cdots \, .
\end{equation}    
We can see that the effect of $Q$ is to slightly reduce the second
order correction to the deflection angle. Therefore, we can conclude
that the sensitivity of future $X$-ray microforlensing experiments to
near-extremal black holes is actually shown in Fig.~\ref{fig:2}.

In summary, evaporation constraints on $f_{\rm PBH}$ can
be substantially altered when moving away from the Schwarzschild
picture. (Other PBH models where Hawking radiation can be slowed down have been recently discussed in~\cite{Calza:2024fzo,Calza:2024xdh}.) Microlensing experiments of $X$-ray pulsars are well positioned to uncover these models.

\section{Conclusions}
\label{sec:4}

We have investigated the potential of searching for bulk black hole dark
matter through microlensing events using future $X$-ray telescopes. We
have shown that for these telescopes the microlensing signal
of a 5D black hole would be indistinguishable from that of a 4D black
hole of the same mass, and this is independent on whether the black
hole is localized on the brane or propagates through the bulk. We have
demonstrated that future instruments observing microlensing events
from $X$-ray pulsars will probe the mass range relevant for bulk
black hole dark matter, including the interesting region in which a
4D PBH all dark matter interpretation has been excluded due to the
non-observation of Hawking radiation.

We end with an observation. PBHs may experience a memory burden
effect, which splits the evaporation process into two distinct phases: semiclassical and quantum~\cite{Dvali:2018xpy,Dvali:2020wft}. If this were the case, then
the minimum black hole mass allowing a PBH all-dark-matter
interpretation would also be
relaxed~\cite{Alexandre:2024nuo,Dvali:2024hsb,Thoss:2024hsr,Balaji:2024hpu}. The quantum decay rate has an additional
suppression factor compared to the Hawking decay rate, which in the
most realistic scenario scales with the inverse of the black hole
entropy. For $d=4$ and a quantum decay rate $\Gamma \sim T_H/S$, the
memory burden effect opens a new window in the mass range $10^9 \alt
M/{\rm g} \alt 10^{10}$~\cite{Thoss:2024hsr}. In contrast to what is
stated in~\cite{Tamta:2024pow}, using (\ref{as}) we argue that due to the
extended nature of the pulsar emission region  $X$-ray microlensing
experiments would be inappropriate to search for signals of the memory burden effect.

\section*{Acknowledgements}

We thank Djuna Croon for permission to reproduce Fig.~\ref{fig:1}.
The work of L.A.A. and K.P.C. is supported by the U.S. National Science
Foundation (NSF Grant PHY-2112527). I.A. is supported by the Second
Century Fund (C2F), Chulalongkorn University.  The work of D.L. is supported by the Origins
Excellence Cluster and by the German-Israel-Project (DIP) on Holography and the Swampland.

\end{document}